\documentclass[12pt]{article}
\usepackage{epsfig}
\usepackage{amsmath}
\usepackage{hhline}
\usepackage{amssymb}
\usepackage{times}
\usepackage{cite}

\newlength{\dinwidth}
\newlength{\dinmargin}
\begin{document}

\newcommand{\ssss}{Schildknecht, Schuler and Surrow}
\newcommand{\mrtt}{Martin, Ryskin and Teubner}
\newcommand{\rcc}{Royen and Cudell}
\newcommand{\ikk}{Ivanov and Kirschner}

%
\newcommand{\s}{\mbox{$s$}}
\newcommand{\ttra}{\mbox{$t$}}
\newcommand{\modt}{\mbox{$|t|$}}
\newcommand{\eminpz}{\mbox{$E-p_z$}}
\newcommand{\eminpzs}{\mbox{$\Sigma(E-p_z)$}}
\newcommand{\rap}{\ensuremath{\eta^*} }
\newcommand{\W}{\mbox{$W$}}
\newcommand{\w}{\mbox{$W$}}
\newcommand{\Q}{\mbox{$Q$}}
\newcommand{\q}{\mbox{$Q$}}
\newcommand{\xB}{\mbox{$x$}}  
\newcommand{\xF}{\mbox{$x_F$}}  
\newcommand{\xg}{\mbox{$x_g$}}  
\newcommand{\xbj}{x}
\newcommand{\xpom}{x_{\PO}}
\newcommand{\y}{\mbox{$y~$}}
\newcommand{\Qsq}{\mbox{$Q^2$}}
\newcommand{\qsq}{\mbox{$Q^2$}}
\newcommand{\kjet}{\mbox{$k_{T\rm{jet}}$}}
\newcommand{\xjet}{\mbox{$x_{\rm{jet}}$}}
\newcommand{\Ejet}{\mbox{$E_{\rm{jet}}$}}
\newcommand{\thjet}{\mbox{$\theta_{\rm{jet}}$}}
\newcommand{\pjet}{\mbox{$p_{T\rm{jet}}$}}
\newcommand{\et}{\mbox{$E_T~$}}
\newcommand{\kt}{\mbox{$k_T~$}}
\newcommand{\ptrans}{\mbox{$p_T~$}}
\newcommand{\pth}{\mbox{$p_T^h~$}}
\newcommand{\pte}{\mbox{$p_T^e~$}}
\newcommand{\ptsq}{\mbox{$p_T^{\star 2}~$}}
\newcommand{\as}{\mbox{$\alpha_s~$}}
\newcommand{\ycut}{\mbox{$y_{\rm cut}~$}}
\newcommand{\gx}{\mbox{$g(x_g,Q^2)$~}}
\newcommand{\xpart}{\mbox{$x_{\rm part~}$}}
\newcommand{\mrsdm}{\mbox{${\rm MRSD}^-~$}}
\newcommand{\mrsdmp}{\mbox{${\rm MRSD}^{-'}~$}}
\newcommand{\mrsdn}{\mbox{${\rm MRSD}^0~$}}
\newcommand{\lambdams}{\mbox{$\Lambda_{\rm \bar{MS}}~$}}
%
%
\newcommand{\gp}{\ensuremath{\gamma}p }
\newcommand{\gammasp}{\ensuremath{\gamma}*p }
\newcommand{\gammap}{\ensuremath{\gamma}p }
\newcommand{\gsp}{\ensuremath{\gamma^*}p }
\newcommand{\dsiget}{\ensuremath{{\rm d}\sigma_{ep}/{\rm d}E_t^*} }
\newcommand{\dsigrap}{\ensuremath{{\rm d}\sigma_{ep}/{\rm d}\eta^*} }
\newcommand{\epem}{\mbox{$e^+e^-$}}
\newcommand{\ep}{\mbox{$ep~$}}
\newcommand{\epl}{\mbox{$e^{+}$}}
\newcommand{\emi}{\mbox{$e^{-}$}}
\newcommand{\epm}{\mbox{$e^{\pm}$}}
\newcommand{\se}{section efficace}
\newcommand{\ses}{sections efficaces}
%
%
\newcommand{\phib}{\mbox{$\varphi$}}
\newcommand{\rh}{\mbox{$\rho$}}
\newcommand{\rhz}{\mbox{$\rh^0$}}
\newcommand{\ph}{\mbox{$\phi$}}
\newcommand{\om}{\mbox{$\omega$}}
\newcommand{\ome}{\mbox{$\omega$}}
\newcommand{\jpsi}{\mbox{$J/\psi$}}
\newcommand{\pipi}{\mbox{$\pi^+\pi^-$}}
\newcommand{\pip}{\mbox{$\pi^+$}}
\newcommand{\pim}{\mbox{$\pi^-$}}
\newcommand{\kk}{\mbox{K^+K^-$}}
\newcommand{\bsl}{\mbox{$b$}}
\newcommand{\alp}{\mbox{$\alpha^\prime$}}
\newcommand{\alpom}{\mbox{$\alpha_{\PO}$}}
\newcommand{\alpomp}{\mbox{$\alpha_{\PO}^\prime$}}
\newcommand{\rzzzz}{\mbox{$r_{00}^{04}$}}
\newcommand{\rzqzz}{\mbox{$r_{00}^{04}$}}
\newcommand{\rzquz}{\mbox{$r_{10}^{04}$}}
\newcommand{\rzqumu}{\mbox{$r_{1-1}^{04}$}}
\newcommand{\ruuu}{\mbox{$r_{11}^{1}$}}
\newcommand{\ruzz}{\mbox{$r_{00}^{1}$}}
\newcommand{\ruuz}{\mbox{$r_{10}^{1}$}}
\newcommand{\ruumu}{\mbox{$r_{1-1}^{1}$}}
\newcommand{\rduz}{\mbox{$r_{10}^{2}$}}
\newcommand{\rdumu}{\mbox{$r_{1-1}^{2}$}}
\newcommand{\rcuu}{\mbox{$r_{11}^{5}$}}
\newcommand{\rczz}{\mbox{$r_{00}^{5}$}}
\newcommand{\rcuz}{\mbox{$r_{10}^{5}$}}
\newcommand{\rcumu}{\mbox{$r_{1-1}^{5}$}}
\newcommand{\rsuz}{\mbox{$r_{10}^{6}$}}
\newcommand{\rsumu}{\mbox{$r_{1-1}^{6}$}}
\newcommand{\rzqik}{\mbox{$r_{ik}^{04}$}}
\newcommand{\rhzik}{\mbox{$\rh_{ik}^{0}$}}
\newcommand{\rhqik}{\mbox{$\rh_{ik}^{4}$}}
\newcommand{\rhaik}{\mbox{$\rh_{ik}^{\alpha}$}}
\newcommand{\rhzzz}{\mbox{$\rh_{00}^{0}$}}
\newcommand{\rhqzz}{\mbox{$\rh_{00}^{4}$}}
\newcommand{\raik}{\mbox{$r_{ik}^{\alpha}$}}
\newcommand{\razz}{\mbox{$r_{00}^{\alpha}$}}
\newcommand{\rauz}{\mbox{$r_{10}^{\alpha}$}}
\newcommand{\raumu}{\mbox{$r_{1-1}^{\alpha}$}}

\newcommand{\R}{\mbox{$R$}}
\newcommand{\rzero}{\mbox{$r_{00}^{04}$}}
\newcommand{\rone}{\mbox{$r_{1-1}^{1}$}}
\newcommand{\costh}{\mbox{$\cos\theta$}}
\newcommand{\cosths}{\mbox{$\cos\theta$}}
\newcommand{\cosp}{\mbox{$\cos\psi$}}
\newcommand{\costop}{\mbox{$\cos(2\psi)$}}
\newcommand{\cosd}{\mbox{$\cos\delta$}}
\newcommand{\cossqp}{\mbox{$\cos^2\psi$}}
\newcommand{\cossqt}{\mbox{$\cos^2\theta^*$}}
\newcommand{\sint}{\mbox{$\sin\theta^*$}}
\newcommand{\sintot}{\mbox{$\sin(2\theta^*)$}}
\newcommand{\sinsqt}{\mbox{$\sin^2\theta^*$}}
\newcommand{\costhst}{\mbox{$\cos\theta^*$}}
\newcommand{\mpipi}{\mbox{$m_{\pi^+\pi^-}$}}
\newcommand{\mkk}{\mbox{$m_{KK}$}}
\newcommand{\mkaka}{\mbox{$m_{K^+K^-}$}}
\newcommand{\mpp}{\mbox{$m_{\pi\pi}$}}       
\newcommand{\mppsq}{\mbox{$m_{\pi\pi}^2$}}   
\newcommand{\mpi}{\mbox{$m_{\pi}$}}          
\newcommand{\mrho}{\mbox{$m_{\rho}$}}        
\newcommand{\mrhosq}{\mbox{$m_{\rho}^2$}}    
\newcommand{\Gmpp}{\mbox{$\Gamma (\mpp)$}}   
\newcommand{\Gmppsq}{\mbox{$\Gamma^2(\mpp)$}}
\newcommand{\Grho}{\mbox{$\Gamma_{\rho}$}}   
\newcommand{\grho}{\mbox{$\Gamma_{\rho}$}}   
\newcommand{\Grhosq}{\mbox{$\Gamma_{\rho}^2$}}   
%
%
\newcommand{\cm}{\mbox{\rm cm}}
\newcommand{\GeV}{\mbox{\rm GeV}}
\newcommand{\gev}{\mbox{\rm GeV}}
\newcommand{\GeVx}{\rm GeV}
\newcommand{\gevx}{\rm GeV}
\newcommand{\GeVc}{\rm GeV/c}
\newcommand{\gevc}{\rm GeV/c}
\newcommand{\MeVc}{\rm MeV/c}
\newcommand{\mevc}{\rm MeV/c}
\newcommand{\MeV}{\mbox{\rm MeV}}
\newcommand{\mev}{\mbox{\rm MeV}}
\newcommand{\MeVx}{\mbox{\rm MeV}}
\newcommand{\mevx}{\mbox{\rm MeV}}
\newcommand{\GeVsq}{\mbox{${\rm GeV}^2$}}
\newcommand{\gevsq}{\mbox{${\rm GeV}^2$}}
\newcommand{\gevsqc}{\mbox{${\rm GeV^2/c^4}$}}
\newcommand{\gevcsq}{\mbox{${\rm GeV/c^2}$}}
\newcommand{\mevcsq}{\mbox{${\rm MeV/c^2}$}}
\newcommand{\GeVsqm}{\mbox{${\rm GeV}^{-2}$}}
\newcommand{\gevsqm}{\mbox{${\rm GeV}^{-2}$}}
\newcommand{\nb}{\mbox{${\rm nb}$}}
\newcommand{\nbinv}{\mbox{${\rm nb^{-1}}$}}
\newcommand{\pbinv}{\mbox{${\rm pb^{-1}}$}}
\newcommand{\mm}{\mbox{$\cdot 10^{-2}$}}
\newcommand{\mmm}{\mbox{$\cdot 10^{-3}$}}
\newcommand{\mmmm}{\mbox{$\cdot 10^{-4}$}}
\newcommand{\degr}{\mbox{$^{\circ}$}}
%
%
\newcommand{\F}{$ F_{2}(x,Q^2)\,$}  
\newcommand{\Fc}{$ F_{2}\,$}    
\newcommand{\XP}{x_{{I\!\!P}/{p}}}       
\newcommand{\TOSS}{x_{{i}/{\PO}}}        
\newcommand{\un}[1]{\mbox{\rm #1}} 
\newcommand{\LO}{Leading Order}
\newcommand{\NLO}{Next to Leading Order}
\newcommand{\ft}{$ F_{2}\,$}
%
%
\newcommand{\mc}{\multicolumn}
\newcommand{\bce}{\begin{center}}
\newcommand{\ece}{\end{center}}
\newcommand{\beq}{\begin{equation}}
\newcommand{\eeq}{\end{equation}}
%
%
\def\lsim{\mathrel{\rlap{\lower4pt\hbox{\hskip1pt$\sim$}}
    \raise1pt\hbox{$<$}}}         
\def\gsim{\mathrel{\rlap{\lower4pt\hbox{\hskip1pt$\sim$}}
    \raise1pt\hbox{$>$}}}         
%
%
\newcommand{\pom}{{I\!\!P}}
\newcommand{\PO}{I\!\!P}
\newcommand{\slowpi}{\pi_{\mathit{slow}}}
\newcommand{\fiidiii}{F_2^{D(3)}}
\newcommand{\fiidiiiarg}{\fiidiii\,(\beta,\,Q^2,\,x)}
\newcommand{\n}{1.19\pm 0.06 (stat.) \pm0.07 (syst.)}
\newcommand{\nz}{1.30\pm 0.08 (stat.)^{+0.08}_{-0.14} (syst.)}
\newcommand{\fiidiiiful}{F_2^{D(4)}\,(\beta,\,Q^2,\,x,\,t)}
\newcommand{\fiipom}{\tilde F_2^D}
\newcommand{\ALPHA}{1.10\pm0.03 (stat.) \pm0.04 (syst.)}
\newcommand{\ALPHAZ}{1.15\pm0.04 (stat.)^{+0.04}_{-0.07} (syst.)}
\newcommand{\fiipomarg}{\fiipom\,(\beta,\,Q^2)}
\newcommand{\pomflux}{f_{\pom / p}}
\newcommand{\nxpom}{1.19\pm 0.06 (stat.) \pm0.07 (syst.)}
\newcommand {\gapprox}
   {\raisebox{-0.7ex}{$\stackrel {\textstyle>}{\sim}$}}
\newcommand {\lapprox}
   {\raisebox{-0.7ex}{$\stackrel {\textstyle<}{\sim}$}}
\newcommand{\pomfluxarg}{f_{\pom / p}\,(x_\pom)}
\newcommand{\dsf}{\mbox{$F_2^{D(3)}$}}
\newcommand{\dsfva}{\mbox{$F_2^{D(3)}(\beta,Q^2,x_{I\!\!P})$}}
\newcommand{\dsfvb}{\mbox{$F_2^{D(3)}(\beta,Q^2,x)$}}
\newcommand{\dsfpom}{$F_2^{I\!\!P}$}
\newcommand{\gap}{\stackrel{>}{\sim}}
\newcommand{\lap}{\stackrel{<}{\sim}}
\newcommand{\fem}{$F_2^{em}$}
\newcommand{\tsnmp}{$\tilde{\sigma}_{NC}(e^{\mp})$}
\newcommand{\tsnm}{$\tilde{\sigma}_{NC}(e^-)$}
\newcommand{\tsnp}{$\tilde{\sigma}_{NC}(e^+)$}
\newcommand{\ssst}{$\star \star \star$}
\newcommand{\sssst}{$\star \star \star \star$}
\newcommand{\tw}{\theta_W}
\newcommand{\sw}{\sin{\theta_W}}
\newcommand{\cw}{\cos{\theta_W}}
\newcommand{\sww}{\sin^2{\theta_W}}
\newcommand{\cww}{\cos^2{\theta_W}}
\newcommand{\trm}{m_{\perp}}
\newcommand{\trp}{p_{\perp}}
\newcommand{\trmm}{m_{\perp}^2}
\newcommand{\trpp}{p_{\perp}^2}
\newcommand{\ev}{\'ev\'enement}
\newcommand{\evs}{\'ev\'enements}
\newcommand{\mdv}{mod\`ele \`a dominance m\'esovectorielle}
\newcommand{\mdmv}{mod\`ele \`a dominance m\'esovectorielle}
\newcommand{\mdm}{mod\`ele \`a dominance m\'esovectorielle}
\newcommand{\idiff}{interaction diffractive}
\newcommand{\idiffs}{interactions diffractives}
\newcommand{\pdmv}{production diffractive de m\'esons vecteurs}
\newcommand{\pdmr}{production diffractive de m\'esons \rh}
\newcommand{\pdmp}{production diffractive de m\'esons \ph}
\newcommand{\pdmo}{production diffractive de m\'esons \om}
\newcommand{\pdm}{production diffractive de m\'esons}
\newcommand{\pdiff}{production diffractive}
\newcommand{\diff}{diffractive}
\newcommand{\produ}{production}
\newcommand{\mv}{m\'eson vecteur}
\newcommand{\mvs}{m\'esons vecteurs}
\newcommand{\me}{m\'eson}
\newcommand{\mr}{m\'eson \rh}
\newcommand{\mph}{m\'eson \ph}
\newcommand{\mo}{m\'eson \om}
\newcommand{\mrs}{m\'esons \rh}
\newcommand{\mps}{m\'esons \ph}
\newcommand{\mos}{m\'esons \om}
\newcommand{\photo}{photoproduction}
\newcommand{\agq}{\`a grand \qsq}
\newcommand{\agqsq}{\`a grand \qsq}
\newcommand{\apq}{\`a petit \qsq}
\newcommand{\apqsq}{\`a petit \qsq}
\newcommand{\de}{d\'etecteur}
%
%
\newcommand{\sqrts}{$\sqrt{s}$}
\newcommand{\Oa}{$O(\alpha_s)$}
\newcommand{\Oaa}{$O(\alpha_s^2)$}
\newcommand{\PT}{p_{\perp}}
\newcommand{\sh}{\hat{s}}
\newcommand{\uh}{\hat{u}}
\newcommand{\ttbs}{\char'134}
\newcommand{\xpomlo}{3\times10^{-4}}
\newcommand{\xpomup}{0.05}
\newcommand{\llq}{$\alpha_s \ln{(\qsq / \Lambda_{QCD}^2)}$}
\newcommand{\llqx}{$\alpha_s \ln{(\qsq / \Lambda_{QCD}^2)} \ln{(1/x)}$}
\newcommand{\llx}{$\alpha_s \ln{(1/x)}$}
%
%
\newcommand{\Brodsky}{Brodsky {\it et al.}}
\newcommand{\FKS}{Frankfurt, Koepf and Strikman}
\newcommand{\Kop}{Kopeliovich {\it et al.}}
\newcommand{\Ginzburg}{Ginzburg {\it et al.}}
\newcommand{\Ryskin}{\mbox{Ryskin}}
\newcommand{\Kaidalov}{Kaidalov {\it et al.}}
%
%
\def\Journal#1#2#3#4{{#1} {\bf #2} (#3) #4}
\def\NCA{Nuovo Cimento}
\def\NIM{Nucl. Instrum. Meth.}
\def\NIMA{Nucl. Instrum. Meth. {\bf A}}
\def\NPB{Nucl. Phys.   {\bf B}}
\def\PLB{Phys. Lett.   {\bf B}}
\def\PRL{Phys. Rev. Lett.}
\def\PRD{Phys. Rev. {\bf D}}
\def\ZPC{Z. Phys. {\bf C}}
\def\EJC{Eur. Phys. J. {\bf C}}
\def\CPC{Comput. Phys. Commun.}

\def\jetp#1#2#3 {JETP Lett. {\bf#1} (#2) #3}

%
%
\newcommand{\clearemptydoublepage}{\newpage{\pagestyle{empty}\cleardoublepage}}
\newcommand{\scaption}[1]{\caption{\protect{\footnotesize  #1}}}
\newcommand{\proc}[2]{\mbox{$ #1 \rightarrow #2 $}}
\newcommand{\average}[1]{\mbox{$ \langle #1 \rangle $}}
\newcommand{\av}[1]{\mbox{$ \langle #1 \rangle $}}

\newcommand{\mphi}{\mbox{$m_{\phi}$}}        
\newcommand{\Gphi}{\mbox{$\Gamma_{\phi}$}}   
\newcommand{\cosdelta}{\mbox{$\cos \delta$}}
\def\mbig#1{\mbox{\rule[-2. mm]{0 mm}{6 mm}#1}}

\begin{titlepage}

\noindent
DESY 00-070  \hspace*{8.5cm} ISSN 0418-9833 \\
May 2000
\vspace{2.0cm}
\begin{center}
\begin{Large}
\boldmath
{\bf  Measurement of elastic electroproduction of $\phi$ mesons at HERA} \\
\unboldmath
\vspace{2cm}
H1 Collaboration
\end{Large}
\end{center}
\vspace{2cm}
 
\begin{abstract}
\noindent
The elastic electroproduction of \ph\ mesons is studied 
at HERA with the H1 detector for photon virtualities 
$1 < Q^2 < 15~{\rm GeV^2}$ and hadronic centre of mass energies
$40 < W < 130$~{\rm GeV}. The \qsq\ and $t$ 
dependences of the cross section are extracted 
($t$  being the square of the four-momentum 
transfer to the target proton).
When plotted as function of (\qsq\ + $M_V^2$) and scaled by the 
appropriate SU(5) quark charge factor, 
the \ph\ meson cross section agrees within errors with the cross sections
of the vector mesons V =  \rh, \om\ and \jpsi.
A detailed analysis is performed of the \ph\ meson polarisation state 
and the ratio of the production 
cross sections for longitudinally and transversely polarised \ph\ mesons 
is determined. A small but significant violation of s-channel helicity
conservation (SCHC) is observed.

\end{abstract}
\vspace{1.5cm}
\begin{center}
To be submitted to {\it Phys. Lett. B}. \\
\end{center}
\end{titlepage}
%
\begin{flushleft}
 C.~Adloff$^{33}$,                
 V.~Andreev$^{24}$,               
 B.~Andrieu$^{27}$,               
 V.~Arkadov$^{35}$,               
 A.~Astvatsatourov$^{35}$,        
 I.~Ayyaz$^{28}$,                 
 A.~Babaev$^{23}$,                
 J.~B\"ahr$^{35}$,                
 P.~Baranov$^{24}$,               
 E.~Barrelet$^{28}$
% ...

\end{flushleft}
\begin{flushleft}
  {\it 
% ...
$ ^{23}$ Institute for Theoretical and Experimental Physics,
          Moscow, Russia \\
 $ ^{24}$ Lebedev Physical Institute, Moscow, Russia$^{e,h}$ \\
% ...
$ ^{27}$ LPNHE, \'{E}cole Polytechnique, IN2P3-CNRS, Palaiseau, France \\
 $ ^{28}$ LPNHE, Universit\'{e}s Paris VI and VII, IN2P3-CNRS,
          Paris, France \\
% ...
$ ^{33}$ Fachbereich Physik, Bergische Universit\"at Gesamthochschule
          Wuppertal, Wuppertal, Germany$^a$ \\
% ...
$ ^{35}$ DESY, Zeuthen, Germany$^a$ \\
% ...
$ ^a$ Supported by the Bundesministerium f\"ur Bildung, Wissenschaft,
        Forschung und Technologie, FRG,
        under contract numbers 7AC17P, 7AC47P, 7DO55P, 7HH17I, 7HH27P,
        7HD17P, 7HD27P, 7KI17I, 6MP17I and 7WT87P \\
% ...
$ ^e$ Supported by the Deutsche Forschungsgemeinschaft \\
% ...
$ ^h$ Supported by Russian Foundation for Basic Research 
       grant no. 96-02-00019 \\
% ...
 }
\end{flushleft}
\newpage
%
%
\section{Introduction}
\noindent
Vector meson production in lepton-proton collisions is a powerful probe
to investigate the nature of diffraction. 
At HERA, because of the wide kinematic ranges in the photon virtuality,
\qsq, and in the hadronic centre of mass energy, \W, the details of the production 
mechanism can be studied. It is also possible to select different vector
mesons, allowing the cross section for different quark types to be studied.
Recent measurements of \rh\ meson electroproduction~\cite{h1_rho_96,zeus_rho_jpsi_hq} 
for high \qsq\ values (\qsq\ $\gsim$ 10 \gevsq) and of \jpsi\ meson photo-- and
electroproduction~\cite{h1_jpsi_gp,h1_jpsi_hq,zeus_rho_jpsi_hq,zeus_jpsi_gp}
show a strong energy dependence of the $\gamma^{\star} p \rightarrow V p$
cross sections. This behaviour indicates that the mass of 
the $c$ quark or a high \qsq\ value provides a hard scale in the interaction,
and we study the elastic cross sections as a function 
of the scale ($Q^2+ M_V^2$), where $M_V$ 
is the mass of the vector meson. 

This paper presents a measurement of elastic \ph\ meson electroproduction
\begin{equation}
e^+ + p \rightarrow e^+ + \ph + p \ \ ;\
\ph \rightarrow K^+ + K^- ,     \label{eq:reac}
\end{equation}
in the \qsq\ range from 1 to 15 \gevsq, and in the \W\ range from 40 to 130 GeV.
The data were obtained with the H1 detector in two running periods when the 
HERA collider was operated with 820~\gev\ protons and 27.5~\gev\ positrons.
A low \qsq\ data set ($1 < \qsq < 5$~\gevsq) with integrated luminosity of 125 \nbinv\
was obtained from a special run in 1995, with the $ep$ interaction
vertex shifted by 70 cm in the outgoing proton beam direction. This results
in a higher acceptance for low $Q^2$ production.
A larger sample of integrated luminosity of 3\pbinv\ with 
$2.5  < \qsq < 15$~\gevsq\ was obtained in 1996
under normal running conditions.
The present measurements provide detailed new information in the region
$1 \lsim  \qsq \lsim 6$~\gevsq\ and they increase the precision of the H1
measurement of \ph\ electroproduction with $\qsq > 6$~\gevsq , which was
first performed using data collected in 1994~\cite{h1_phi_94}.
They are compared to results of the ZEUS experiment in 
photoproduction~\cite{zeus_phi_gp} and at $\qsq > 7$~\gevsq~\cite{zeus_phi_hq}.
The elastic \ph\ meson cross section is also compared to elastic
\rh~\cite{h1_rho_96,zeus_rho_jpsi_hq,zeus_rho_gp},
\om~\cite{zeus_ome_gp},
\jpsi~\cite{zeus_rho_jpsi_hq,h1_jpsi_gp,zeus_jpsi_gp,h1_jpsi_hq} and
$\Upsilon$~\cite{h1_ups_gp,zeus_ups_gp} 
meson production results from H1 and ZEUS.


The event selection and the $K^+K^-$ mass distribution is presented
in section~\ref{sec:signal}. In section~\ref{sec:rat}, the elastic \ph\ cross section
is presented as a function of \qsq\ and \W. 
In order to minimise the uncertainties, the cross section is measured as a ratio 
to elastic \rh\ production, and the absolute elastic \ph\ cross section is then extracted
using the results for $\rho$ production from~\cite{h1_rho_96}. 
A compilation of the \rh, \om, \ph, 
\jpsi, and $\Upsilon$ cross sections is presented as a function of (\qsq\ + $M_V^2$). 
The $t$ dependence of the elastic \ph\ cross section is analysed in
section~\ref{sec:t}. A detailed analysis of the photon and \ph\ meson 
polarisations is performed in  section~\ref{sec:pol} and the 15 spin density matrix
elements are extracted. The ratio $R$ of the longitudinal to transverse \ph\ 
cross sections is obtained as a function of \qsq.  A compilation of the measurements of
$R$ for elastic \rh, \ph\ and \jpsi\ meson production is presented as a 
function of \qsq/$M_V^2$. 

The present analysis uses to a large extent the techniques described
in the H1 publication on elastic \rh\ production~\cite{h1_rho_96}. 


\section{Data selection} \label{sec:signal}
Elastic \ph\ electroproduction events are selected on the basis of their topology
in the H1 detector\footnote{
  A detailed description of the H1 detector can be found in~\protect\cite{h1det}.}.
They must have a positron candidate and two oppositely charged hadron
candidates, originating from a vertex situated in the nominal $e^+p$ interaction 
region, with K$^+$K$^-$ invariant mass in the range from 1.00 to 1.04~GeV.
The scattered positron is identified as an electromagnetic cluster of energy 
larger than 15 GeV detected in the H1 backward electromagnetic calorimeter
SPACAL~\cite{spacal}\footnote{
  H1 uses a right-handed coordinate system with the $z$ axis taken along the
  beam direction, the $+z$ direction being that of the outgoing proton beam. The
  $x$ axis points towards the centre of the HERA ring.}.
The two hadron candidates are recognised as 
tracks of opposite signs, with a
momentum transverse to the beam direction larger than 100 \mev,
reconstructed in the H1 central tracking detector with a polar
angle in the range from 20$^\circ$ to 160$^\circ$. The 
vertex must lie within 30 cm along the beam axis
from the nominal interaction point. The nature of the hadrons 
is not explicitly identified.
Their charge and momentum are measured in the central part of the detector
by means of a uniform 1.15 T magnetic field.
No other activity must be observed in the detector since
the scattered proton remains in the beam pipe and is not 
detected because of the small momentum transfer to the
target in diffractive interactions. Events were therefore rejected if there
were signals in the forward part of the detector 
(forward muon and forward proton tagger detectors)
and if there were clusters in the liquid argon calorimeter 
with an energy above 0.5~GeV not associated with
the hadron candidates.
To reduce effects of QED radiative corrections, the selected events 
have to satisfy $\sum_{e,h} (E-p_z)$ $>$ 45 \gev.

The \qsq\ variable is reconstructed from the incident electron beam
energy and the polar angles of the positron and of the \ph\ meson candidates
(double angle method~\cite{double_angle}). The \W\ variable is 
reconstructed using in addition the energy and the longitudinal 
momentum of the \ph\ meson candidate.

The variable $t$ is the square of the four-momentum transfer to the target proton.
At HERA energies, to a very good precision, the absolute value of $t$ 
is equal to the square of the transverse momentum of the outgoing proton.
The latter is computed, under the assumption that the selected event corresponds
to reaction (\ref{eq:reac}), as the square of the vector sum of the transverse 
momenta of the \ph\ meson candidate and of the scattered positron.
Events with \modt\ $<$ 0.5 \gevsq\ are selected in order to reduce the remaining
production of proton dissociation events which have a flatter $t$ distribution, 
and to suppress the production of hadron systems of which the \ph\ is only part and in 
which the remaining particles were not detected.

\begin{figure}[tp]
\begin{center}
\setlength{\unitlength}{1.0cm}
\begin{picture}(17.5,17.5)
\put(-0.5,0.0){\epsfig{file=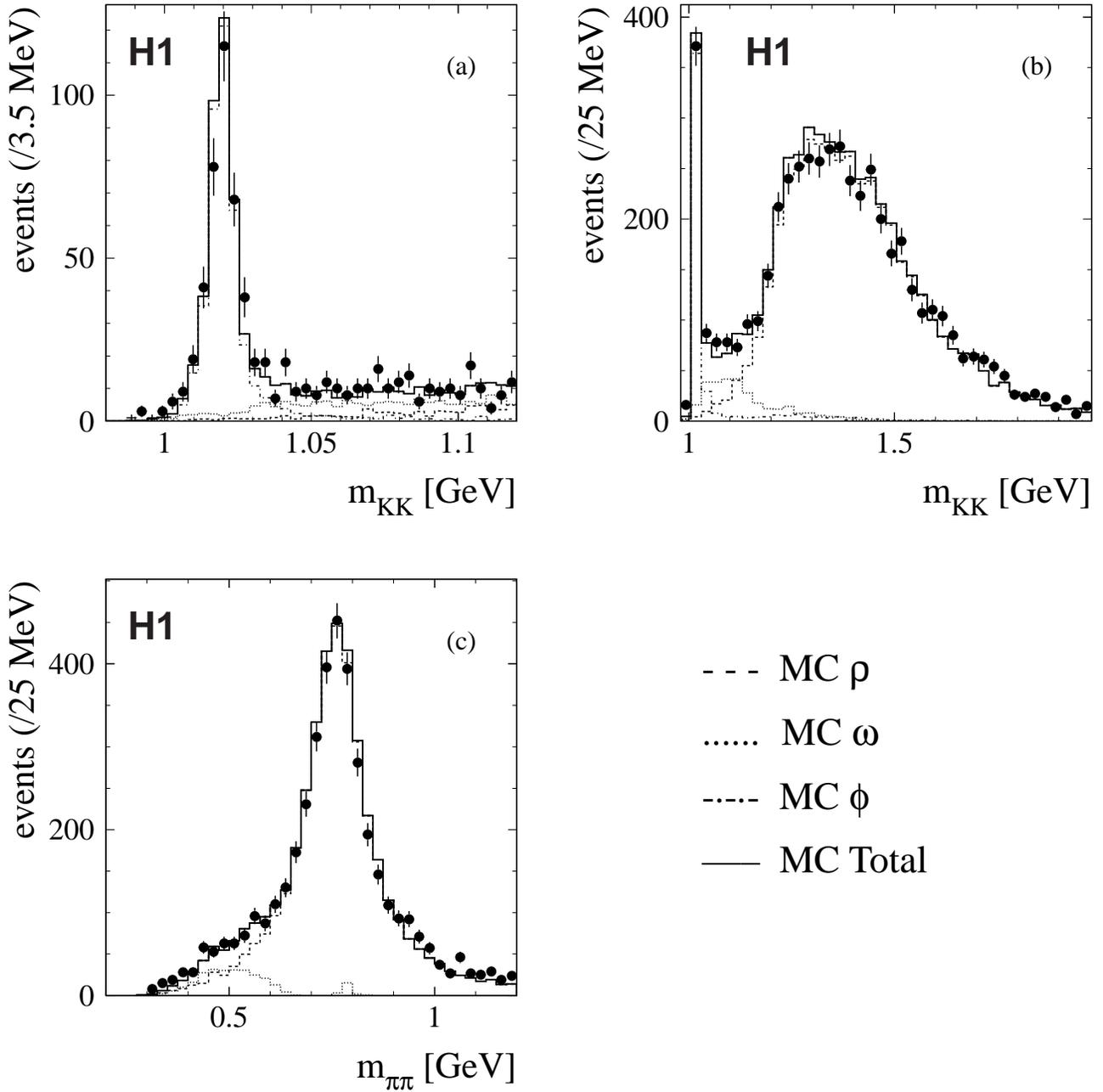}}%
\end{picture}
\caption{(a) and (b): distribution of \mkk\ for \mkk\ $<$ 1.12 GeV
   and \mkk\ $<$ 2.00 GeV, respectively; (c) distribution of \mpp\ after 
   removing the \ph\ signal (\mkk\ $<$ 1.04 GeV). The points
   represent the data and the full lines the prediction of a simulation,
   which includes
   contributions from elastic \rh\ (dashed lines), \om\ (dotted lines) and
   \ph\ (dash-dotted lines) meson production.
   The errors on the data points are statistical only.}
\label{fig:signal}
\end{center}
\end{figure}

The distribution of \mkk, the two particle invariant mass computed under the
assumption that the hadron candidates are kaons,
is presented in Fig.~\ref{fig:signal}a and Fig.~\ref{fig:signal}b 
for \mkk\ $<$ 1.12 GeV and for \mkk\ $<$ 2.00 GeV, respectively. 
A clear \ph\ signal is observed in the data, with 
424 events in the range 1.00 $<$ \mkk\ $<$ 1.04 GeV.

The main backgrounds to reaction (\ref{eq:reac}) are due to diffractive \ph\ events
in which the proton is excited into a system of higher mass which 
subsequently dissociates, and to the elastic production of \rh\ and 
\om\ vector mesons. The other backgrounds (other \ph\ decay channels, higher
mass resonances or non resonant production) are estimated to be less than a few
percent.
The fraction of proton dissociation background is assumed to 
be the same for \ph\ as for \rh\ meson production and is taken to be 
11$\pm$5~\% as in~\cite{h1_rho_96}. The background due to \rh\ and \om\
production is estimated using the DIFFVM simulation~\cite{diffvm}.
The DIFFVM Monte Carlo simulation program is based on Regge theory and on 
the vector meson dominance model. The \rh\ and \om\ backgrounds 
are normalised to the \mpp\ distribution observed in the data, where
\mpp\ is the invariant mass computed under the pion hypothesis for the hadron 
candidates. This is shown in Fig.~\ref{fig:signal}c after the \ph\ signal 
has been removed by selecting \mkk\ $>$ 1.04 GeV. The background under the \ph\ peak
from \rh\ and \om\ meson production is \qsq\ dependent and varies from 
15~\% to 4~\%.
For the full sample (2.5 $<$ \qsq\ $<$ 15 \gevsq) this 
background is found to be  9$\pm$5~\%.

The data are corrected for acceptances, efficiencies and detector 
resolution effects using the DIFFVM Monte Carlo simulation. 
The response of the H1 detector is fully simulated.
%

\boldmath
\section{Elastic cross section} \label{sec:rat}
\unboldmath

The elastic $\phi$ meson production cross section is obtained
by first measuring the ratio of the $\phi$ to $\rho$ cross sections
and then using the $\rho$ cross sections which were precisely
measured as described in~\cite{h1_rho_96}.
In the ratio, several uncertainties
cancel, most notably the luminosity uncertainty, the contribution of the 
proton dissociation background
and the trigger efficiency, which is very similar for both data
samples since it is mostly based on the positron detection in the SPACAL.
The remaining corrections account for the mass selection range and the 
differences in acceptances and $t$ distribution. The correction for the
accepted mass range (0.6 $<$ \mpp\ $<$ 1.1~GeV) in the \rh\ sample is
1.16 $\pm$ 0.02 \mbig{$^{+0.05}_{-0.00}$}~\cite{h1_rho_96}; the correction
for the mass range (1.00 $<$ \mkk\ $<$ 1.04~GeV) in the \ph\ sample is
estimated to be 1.03 $\pm$ 0.01 using the DIFFVM simulation.
In both samples, hadron tracks must be detected in the central tracker
with $20^\circ < \theta < 160^\circ$. Differences in the 
acceptances for the two samples, due to the different decay hadron 
and vector meson 
masses, are estimated as a function of \qsq, \W\ and 
\ttra\ using the DIFFVM simulations for \rh\ and \ph\ production.
Differences in the detector efficiency for pions and kaons are taken into
account in the detector simulation. 
Finally the correction for events with \modt\ $>$ 0.5 \gevsq\ in the
\rh\ and \ph\ samples is estimated by assuming an exponentially falling
\modt\ distribution, using recent measurements for the slope
parameter of the 
exponential~\cite{h1_rho_96,zeus_rho_jpsi_hq,h1_phi_94,zeus_phi_hq,zeus_phi_gp}. 
The correction factor on the \ph/\rh\ cross section ratio is 1.03 $\pm$ 0.02, independent of \qsq.
The branching ratios of 0.49 and 1.0 were used for the decays \ph\ $\rightarrow$ $K^+K^-$ 
and \rh\ $\rightarrow$ $\pi^+\pi^-$ respectively.

Systematic errors on the measurement of the cross section ratio
are estimated by varying all corrections within the errors.
In addition, in both the \rh\ and \ph\ simulations 
the cross section dependence on \qsq, \W, $t$ and the vector meson angular decay 
distributions were varied by amounts allowed by the present and most recent 
measurements~\cite{h1_rho_96,zeus_rho_jpsi_hq,h1_phi_94,zeus_phi_hq,zeus_phi_gp}.
 
\begin{figure}[t]
\begin{center}
\setlength{\unitlength}{1.0cm}
\begin{picture}(12.0,9.0)
\put(0.0,-1.5){\epsfig{file=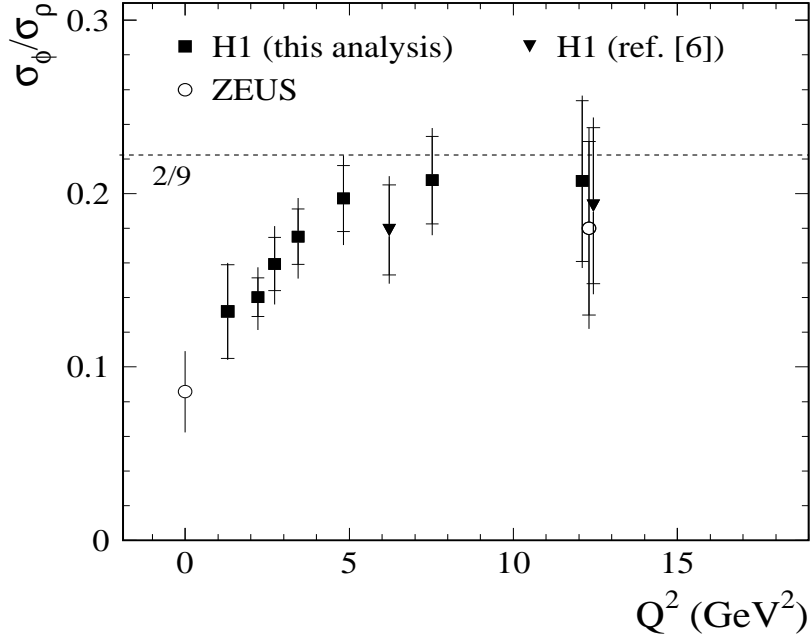,%
            height=11.0cm,width=12.cm}}
\end{picture}
\caption{Ratio of the cross sections for elastic \ph\ and \rh\ production,
    as a function of \qsq, together with 
    previous H1~\protect\cite{h1_phi_94} and 
    ZEUS~\protect\cite{zeus_phi_hq,zeus_phi_gp} measurements.
    The inner error bars are statistical and the full error bars include 
    the systematic errors added in quadrature.
    The dashed line corresponds to the ratio $2/9$.}
\label{fig:ratio}
\end{center}
\end{figure}

The \qsq\ dependence of the \ph\ to \rh\ elastic cross section ratio is
presented in Fig.~\ref{fig:ratio} together with 
previous H1~\cite{h1_phi_94} and 
ZEUS~\cite{zeus_phi_hq,zeus_phi_gp} results.
The values of the ratio are given in Table~\ref{table:values}.
The present measurements confirm the significant rise of the 
cross section ratio with \qsq. As \qsq\ increases, the HERA cross 
section ratios approach the value  $2 / 9$ expected from quark charge counting and 
SU(5). It should be noted that calculations based on 
perturbative QCD predict that the cross section ratio should exceed 
this value at very large \qsq~\cite{fks}.
The \w\ dependence of the \ph\ to \rh\ elastic cross section ratio is
measured in the range 40 $<$ W $<$~130 GeV and is observed to be constant, 
within the experimental uncertainties.


To extract the $\gamma^{\star}p \rightarrow \ph p$ cross section, the measurement of the 
$\phi/\rho$ cross section ratio is multiplied by the $\gamma^{\star}p \rightarrow \rh p$ 
cross section calculated from the fit in~\cite{h1_rho_96}. The values are given in 
Table~\ref{table:values}. The systematic errors on the \ph\ cross section measurement 
include the systematic errors on the ratio of \ph\ to \rh\ cross sections, as well as
an 8.4 \% contribution coming from the parametrisation error in
the fit of the \rh\ cross section (see ~\cite{h1_rho_96}), added in quadrature.

\begin{figure}[p]
\setlength{\unitlength}{1.0cm}
\begin{center}
\begin{picture}(16.0,16.0)
\put(0.0,0.){\epsfig{file=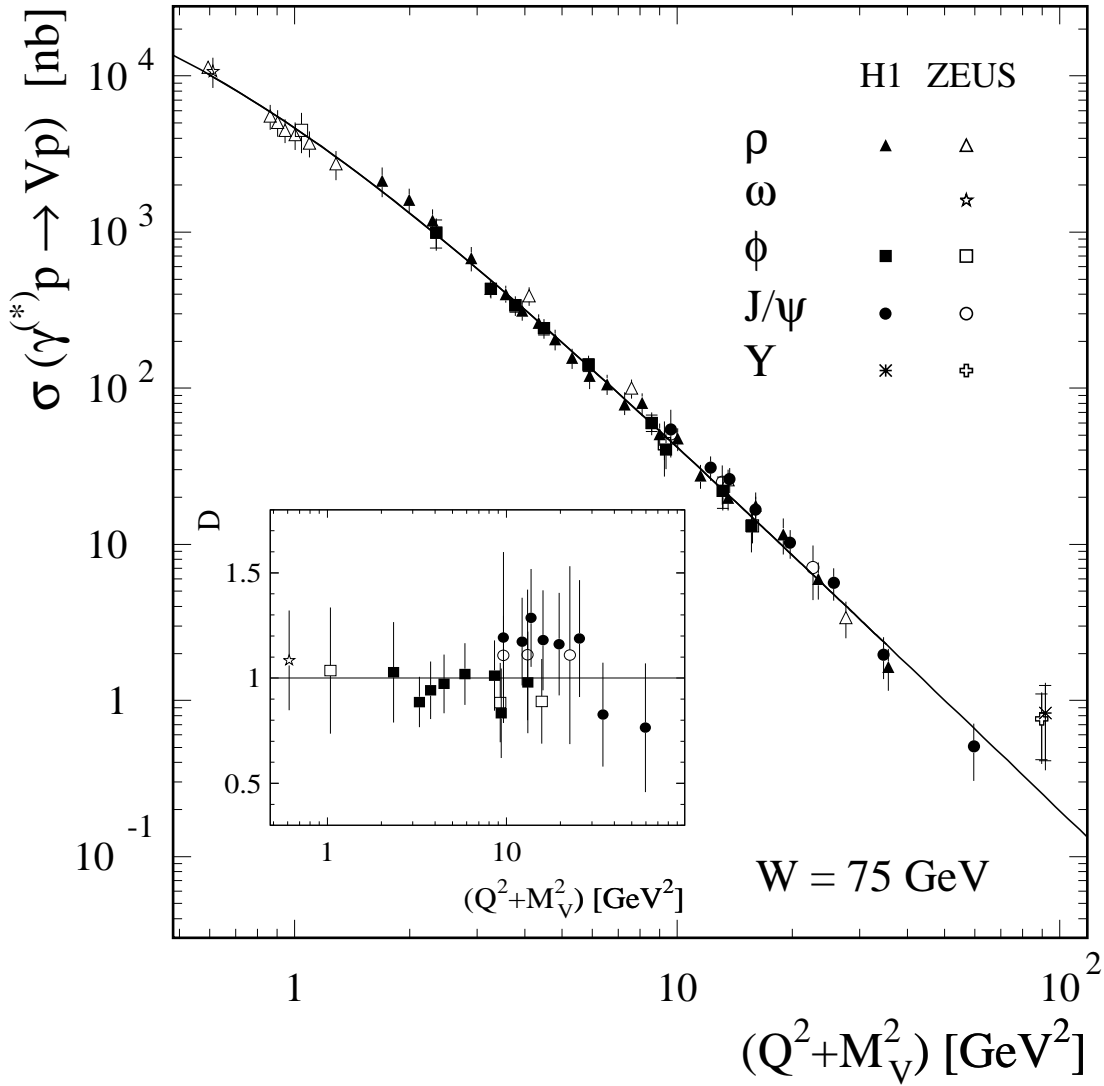}}%
\end{picture}
    \caption{H1 and ZEUS 
     measurements~\protect\cite{h1_rho_96,zeus_rho_jpsi_hq,
     h1_jpsi_gp,h1_jpsi_hq,zeus_jpsi_gp, h1_phi_94,zeus_phi_gp,
     zeus_phi_hq,zeus_rho_gp,zeus_ome_gp,h1_ups_gp,zeus_ups_gp}
     of the total cross sections
     $\sigma (\gamma^{\star}p \rightarrow V p)$ as a function of (\qsq\ + $M^2_V$)
     for elastic \rh, \om, \ph, \jpsi\ and $\Upsilon$ meson production, at
     the fixed value $W$ = 75 GeV. The cross sections were scaled by SU(5) factors, 
     according to the quark charge content of the vector meson. 
     For the error bars statistical and systematic errors have been added in quadrature.
     The curve corresponds to a fit to
     the H1 and ZEUS \rh\ data, and 
     the ratio $D$ of the scaled $\omega$, \ph\ and \jpsi\ cross sections to this
     parametrisation is presented in the insert.}
\label{fig:q2_dep}
\end{center}
\end{figure}

In Fig.~\ref{fig:q2_dep}, the cross section for the elastic production of 
\ph\ mesons (full squares) is presented together with other vector mesons $V$
and for various 
values of \qsq, as a function of the variable (\qsq\ + $M_V^2$).
The data in Fig.~\ref{fig:q2_dep} compile the HERA 
measurements~\cite{h1_rho_96,zeus_rho_jpsi_hq,h1_jpsi_gp,h1_jpsi_hq,zeus_jpsi_gp,
h1_phi_94,zeus_phi_gp,zeus_phi_hq,zeus_rho_gp,zeus_ome_gp,h1_ups_gp,zeus_ups_gp}
of the $\gamma^{\star}p \rightarrow V p$
cross sections (see also~\cite{bar_tel_aviv}). 
The cross sections were scaled by SU(5) factors, according to the
quark charge content of the vector meson, which amount to 1 for the \rh,
9 for the \om, 9/2 for the \ph, 9/8 for the \jpsi\ and 9/2 for
the $\Upsilon$ meson. 
The cross sections are measured at \W\ = 75 GeV, or are moved to that value
according to the parametrisation $\sigma \propto W^\delta$, using the
$\delta$ value measured by the corresponding experiment.
The ZEUS \rh\ and \ph\ cross sections were corrected ($\lsim$ 7 \%)
for the unmeasured signal with $|t|$ $>$ 0.5 (or 0.6) \gevsq\ by assuming a 
simple exponential fall
of d$\sigma/dt \propto e^{bt}$. In this procedure the observed \qsq\ dependence of the
$b$ slope was taken into account.

Within the experimental errors, the total cross sections for vector meson production,
including the SU(5) normalisation factors, appear to lie on a
universal curve when plotted as a function of the scale $(Q^2+M^2_V)$,
except possibly for the $\Upsilon$ photoproduction\footnote
   {The cross sections $\sigma (\gamma p \rightarrow \Upsilon({\rm 1S}) p)$ measured by
    H1 and ZEUS at \W\ = 143 and 120 GeV respectively~\cite{h1_ups_gp,zeus_ups_gp}, 
    were moved to the value
    \W\ = 75 GeV using the parametrisation $\sigma \propto W^\delta$, with $\delta$ = 1.7.
    This high value of the parameter $\delta$ comes from the prediction
    of~\protect\cite{martin}. Note that if the value $\delta$ = 0.8 is used (a value measured
    in case of \jpsi\ photoproduction), the cross sections increase  
    by a factor 1.5 for ZEUS and 1.8 for H1.}.
A fit performed on the H1 and ZEUS \rh\ data using the parametrisation
$\sigma = a_1 (Q^2 + M^2_V + a_2)^{a_3}$, with 
$a_1$ = 10689 $\pm$ 165~nb,
$a_2$ = 0.42 $\pm$ 0.09~\gevsq\ and
$a_3$ = -- 2.37 $\pm$ 0.10 ($\chi^2/ndf$ = 0.67) is shown
as the curve in Fig.~\ref{fig:q2_dep}.
The ratio of the \om, \ph\ and \jpsi\ cross sections to this
parametrisation is presented in the insert of Fig.~\ref{fig:q2_dep}.
Note that the universal $(Q^2 + M^2_V)$ dependence
is for the total cross section measurements only. 
The separate behaviour of the longitudinal
and transverse cross sections is described in ref.~\cite{bar_tel_aviv}.

\begin{table}[tbh]
\begin{center}
\begin{tabular}{|c|c|c|}
\hline
$Q^2$ (GeV$^2$)   &  $\sigma$(\ph)/$\sigma$(\rh)  &  $\sigma (\gamma^\star p \rightarrow \ph p$) (nb) \\
\hline
1.3   & 0.132 $\pm$ 0.027 $\pm$ 0.008               & 220 $\pm$ 45 $\pm$ 24  \\
2.22  & 0.140 $\pm$ 0.011 \mbig{$^{+0.009}_{-0.011}$} &  96.3 $\pm$ 7.6 $\pm$ 10.6 \\
2.73  & 0.159 $\pm$ 0.015 \mbig{$^{+0.012}_{-0.015}$} &  75.3 $\pm$ 7.1 $\pm$ 8.3  \\
3.44  & 0.175 $\pm$ 0.016 \mbig{$^{+0.012}_{-0.015}$} &  53.7 $\pm$ 4.9 $\pm$ 5.9  \\
4.82  & 0.197 $\pm$ 0.019 \mbig{$^{+0.013}_{-0.016}$} &  31.3 $\pm$ 3.0 $\pm$ 3.4  \\
7.53  & 0.208 $\pm$ 0.025 \mbig{$^{+0.013}_{-0.017}$} &  13.3 $\pm$ 1.6 $\pm$ 1.5  \\
12.1  & 0.207 $\pm$ 0.046 \mbig{$^{+0.013}_{-0.017}$} &  4.9  $\pm$ 1.1 $\pm$ 0.5  \\
\hline
\end{tabular}
\end{center}
 \caption{Ratio of the cross sections for elastic \ph\ and \rh\ production 
    and the elastic \ph\ meson cross sections $\sigma (\gamma^\star p \rightarrow \ph p)$,
    for seven \qsq\ values.
    The cross sections are given for \W\ = 75 GeV.
    The first error represents the statistical error and the second the
    systematic error.}
\label{table:values}
\end{table}

\boldmath
\section{Dependence on $t$}  \label{sec:t}
\unboldmath

In this section and the following one, the elastic 
\ph\ meson production is studied using the \ph\ sample defined 
above, with the additional selection: the centre of gravity of
the scattered positron cluster was required to lie outside the 
innermost part of the SPACAL calorimeter $-16 < x < 8$~cm 
and $-8 < y < 16$~cm in order to obtain good ($>$ 95 \%) and 
uniform trigger efficiency. The number of elastic \ph\ candidates 
is then reduced to 221 events for 2.5 $<$ \qsq\ $<$ 15 \gevsq.

\begin{figure}[bt]
\begin{center}
\setlength{\unitlength}{1.0cm}
\begin{picture}(9.5,10.0)
\put(-0.4,0.0){\epsfig{file=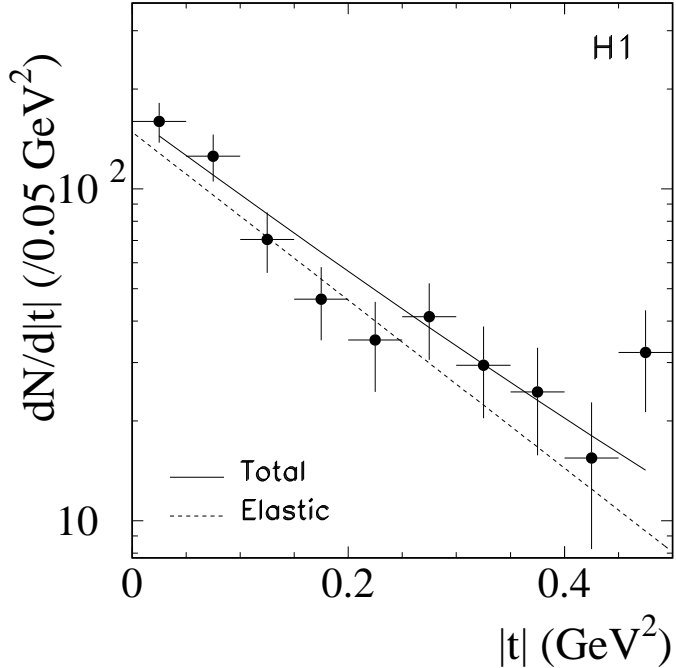,%
            height=10cm}}
\end{picture}
\caption{Corrected $t$ distributions for elastic \ph\ meson production, in the range
2.5 $<$ \qsq\ $<$ 15~\gevsq.
The full curve presents the result of the fit to the sum of
three exponentials (see text), the dotted line showing the elastic contribution. 
The errors on the data points are statistical only.}
\label{fig:t_dep}
\end{center}
\end{figure}

The measured  \modt\ dependence is shown in Fig.~\ref{fig:t_dep} and the
characteristic falling exponential distribution is observed.
To take into account the contribution of different backgrounds, the \modt\ distribution 
is fitted by the sum of three exponentials corresponding
to the elastic \ph\ component, the diffractive \ph\ component with proton dissociation
and the $\omega$ and \rh\ production.
The elastic \ph\ component is fitted with a free normalisation and a free 
slope parameter $b$, whereas the other contributions are fixed to their
calculated values.
The contribution of diffractive \ph\ events with proton dissociation 
of $11 \pm 5\,\%$ of the elastic signal and a slope parameter
of $2.5 \pm 1.0$~\gevsqm, was taken from~\cite{h1_rho_96}.
The $\omega$ and $\rh$ background contributions, 
amounting to $9 \pm 5\,\%$ of the signal (see section~2),
have an effective slope parameter $b$ = 2.9 $\pm$ 0.6~\gevsqm, 
computed using the DIFFVM simulation.

The fitted exponential slope parameter for elastic \ph\ events is found
to be $ b = $5.8$\pm$0.5 (stat.) $\pm$0.6 (syst.) \gevsqm, 
for an average \qsq\ value of 4.5 \gevsq\ and \average{W}=75~\gev. 
The systematic error is computed by varying the amounts of the background
contributions and their slopes within the quoted errors, and by varying 
the binning and the limits of the fit. The effect
of the QED radiative corrections on the $b$ measurement
is estimated using the simulation DIFFVM including
a HERACLES~\cite{heracles} interface, and is found to decrease 
the value of the $b$ measurement by 0.13~\gevsqm\ (the $b$ value given above
is not corrected for this effect). 
This result can be compared with
other measurements, $b = 7.3 \pm 1.0 \pm 0.8$~\gevsqm\ 
in photoproduction~\cite{zeus_phi_gp} and 
$b = 5.2 \pm 1.6 \pm 1.0$~\gevsqm\ for $\langle Q^2 \rangle $ = 10~\gevsq~\cite{h1_phi_94}.
The data are consistent with a decrease of the slope parameter 
as \qsq\ increases; this would be expected from the decrease of the 
transverse size of the virtual photon.

The value of the $b$ slope parameter is in agreement within the errors
with the one obtained in elastic \rh\ meson production: $b$ = 5.5 $\pm$ 0.5 (stat.)
\mbig{$^{+0.5}_{-0.2}$} (syst.)~\gevsqm, at \qsq\ = 4.8 \gevsq~\cite{h1_rho_96}.
%

\boldmath
\section{Polarisation studies}  \label{sec:pol}
\unboldmath
The study of the angular distributions of the production and decay of the
\ph~meson provides information on the photon and \ph\ meson polarisation states.
In the helicity system~\cite{schilling-wolf}, three angles are defined as follows.
The angle $\Phi$, defined in the hadronic centre of mass system (cms), is the
azimuthal angle between the electron scattering plane and the plane containing
the \ph\ meson and the scattered proton.
The \ph\ meson decay is described by the polar angle $\theta$ and the azimuthal
angle \phib\ of the positive kaon in the $K^+K^-$ rest frame, with the
quantisation axis taken as the direction opposite to that of the outgoing
proton in the hadronic cms (the so called helicity frame).
Details of the kinematics and the mathematical formalism can be found
in~\cite{schilling-wolf} and~\cite{h1_rho_96}.
The normalised angular decay distribution $F$(\cosths, \phib, $\Phi$) is
expressed as a function 
of 15 spin density matrix elements corresponding to 
different bilinear combinations of the helicity amplitudes
$T_{\lambda_{\phi}, \lambda_{\gamma}}$, where
$\lambda_{\phi}$ and $\lambda_{\gamma}$ are the helicities of the \ph\
meson and of the photon, respectively. In the case of $s$-channel helicity
conservation (SCHC), the helicities of the \ph\ meson and the photon
are equal, only the amplitudes $T_{00}$, $T_{11}$, and $T_{-1-1}$ are 
different from zero and 10 of the 15 matrix elements are zero.

\begin{figure}[thbp]
\setlength{\unitlength}{1.0cm}
\begin{picture}(16.0,19.0)
\put(0.0,0.0){\epsfig{file=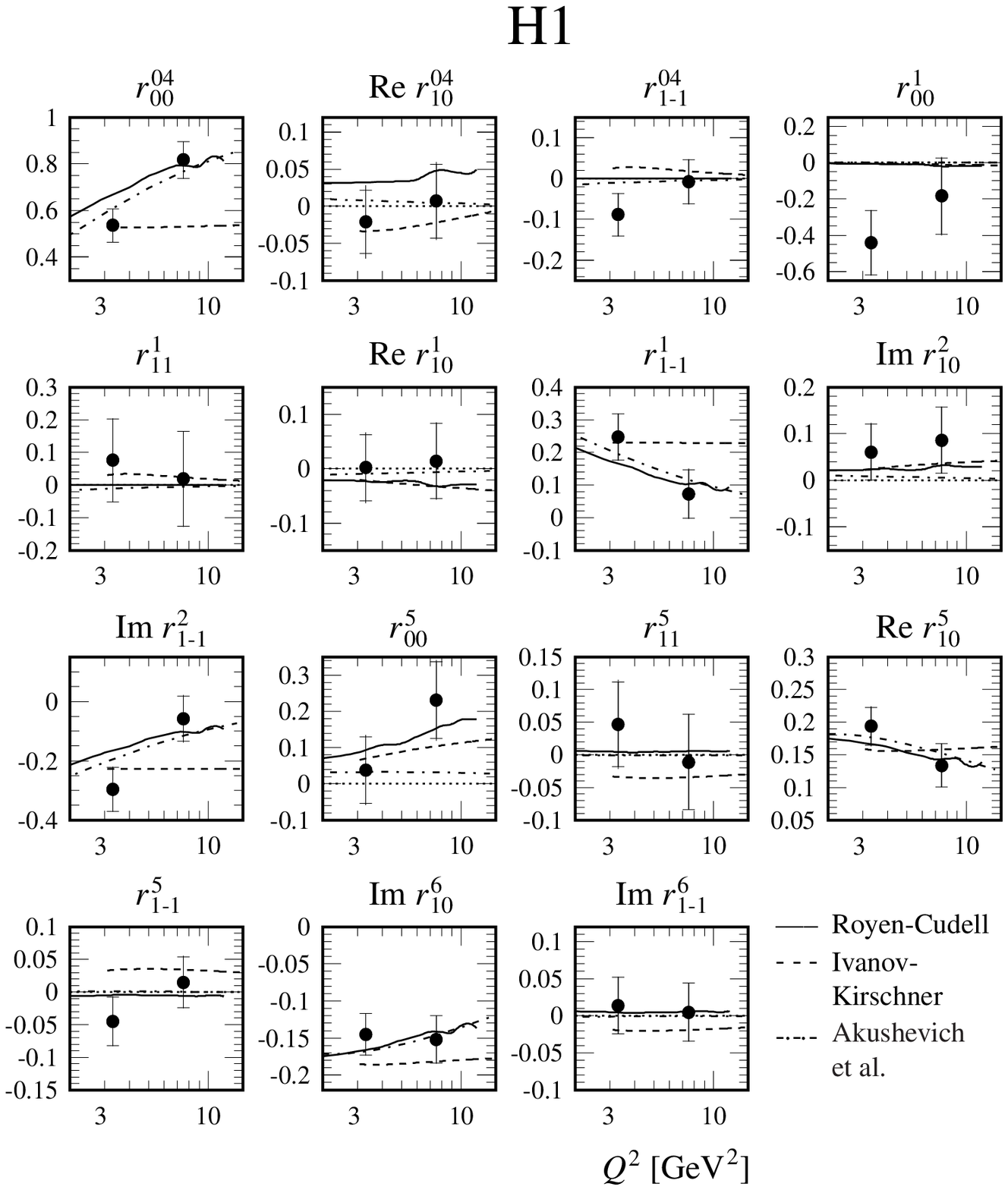}}%
\end{picture}
\caption{The full set of spin density matrix elements for elastic electroproduction
  of \ph\ mesons, for two ranges of \qsq.
  The inner bars are statistical, the full error bars include 
  the systematic errors added in quadrature.
  Where indicated the dotted lines show the expectation of zero for
  $s$-channel helicity conservation (SCHC). The five elements predicted to be non-zero
  under SCHC are  \rzqzz, \ruumu, Im \rdumu, Re \rcuz\ and Im \rsuz.
  The full, dashed and dash-dotted lines represent respectively predictions
  of the models of Royen and Cudell~\protect\cite{isa},
  Ivanov and Kirschner~\protect\cite{ivanov}
  and Akushevich, Ivanov and Nikolaev~\protect\cite{kolya}. }
\label{fig:matqsq}
\end{figure}

The matrix elements are measured using projections of the decay angular
distribution onto orthogonal trigonometric functions of the angles
$\theta$, \phib\ and $\Phi$~\cite{schilling-wolf}.
The results are presented in Fig.~\ref{fig:matqsq} in two \qsq\ bins:
$2.5 < Q^2 < 4.5 $ \gevsq\ and  $4.5 < Q^2 < 15 $ \gevsq.
In Fig.~\ref{fig:matqsq}, the results are not corrected for 
the small effects due to proton dissociation, \om\ and \rh\ production 
backgrounds and radiative effects.

%

The matrix elements generally follow the SCHC predictions, except 
for the elements \ruzz\ and \rczz, which may indicate a small 
violation of SCHC. The matrix element \rczz\ is proportional to 
the product $T^*_{00} T_{01}$ of helicity amplitudes,
the dominant SCHC violating amplitude
being $T_{01}$ ($\lambda_{\phi}$ = 0 and $\lambda_{\gamma}$ = 1).

Predictions from recent models based on perturbative QCD~\cite{isa,ivanov,kolya}
are compared to the measurement of the 15 matrix elements.
The models are expected to be 
valid at high \qsq\ (providing a scale for the perturbative expansion)
and at high energy: $W^2 \gg \qsq \gg \Lambda^2_{{\rm QCD}}$.
The \ph\ meson production is factorised, in the proton rest
frame, into three parts involving different time scales:
the fluctuation of the photon into a $q\bar{q}$ state,  at a large
distance from the target, the hard scattering of the
$q \bar{q}$ pair with the proton, modelled as two-gluon exchange,
and the $q \bar{q}$ pair recombination into a \ph\ meson wave.
The amplitudes are computed separately for the different helicities of the
photon and the \ph\ meson. In models~\cite{ivanov,kolya}, 
the gluon density in the proton is used for the computation of the hard
scattering amplitude.
Differences between the models are related to the way of
introducing quark off-shellness and Fermi motion. All models describe 
the data relatively well, predicting in particular a non-zero value 
for the \rczz\ matrix element (see Fig.~\ref{fig:matqsq}). The model~\cite{ivanov}
gives a poorer description of the \qsq\ dependence of the \rzqzz, \ruumu\ and Im~\rdumu\
matrix elements, which are correlated, than the models of~\cite{isa,kolya}. 

Another way to study the violation of SCHC is to measure the $\Phi$ 
angular distribution:
$F(\Phi) \propto  1 - \varepsilon\ \cos2\Phi\ (2  \ruuu + \ruzz) +
   \sqrt{2\varepsilon (1+\varepsilon)} \cos \Phi\ (2  \rcuu + \rczz)$,
where $\varepsilon$ is the polarisation parameter of the virtual photon.
In the case of SCHC, this distribution is predicted to be uniform, the 
matrix elements \ruuu, \ruzz, \rcuu\ and \rczz\ being zero.

\begin{figure}[p]
\begin{center}
\setlength{\unitlength}{1.0cm}
\begin{picture}(16.0,8.0)
\put(-0.4,0.0){\epsfig{file=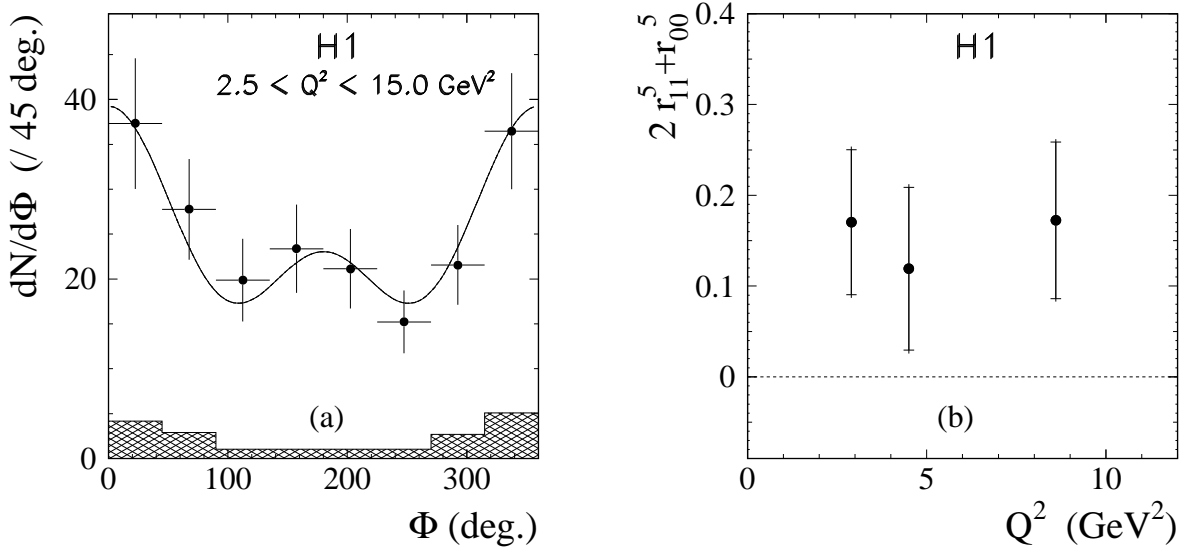}}%
\end{picture}
\caption{ (a) Background corrected $\Phi$ angle distribution, 
    the curve represents the result of a fit to
    $F(\Phi)$ (see text) and the hashed area the subtracted background,
    the errors on the data points are statistical only;
    (b) value of the combination of the matrix elements
    $(2 \rcuu + \rczz)$ for three \qsq\ bins, after background correction and 
    QED radiative effects taken into account, the dotted line shows the zero
    prediction from SCHC, the inner error bars are statistical and the full 
    error bars include the systematic errors added in quadrature.}
\label{fig:phi2}
\end{center}
\end{figure}

The $\Phi$ distribution for the elastic \ph\ meson production is 
presented in Fig.~\ref{fig:phi2}a. The distribution is corrected 
for the presence of \rh\ and \om\ backgrounds (hashed area).
The result of the fit to the function $F(\Phi)$ is given as the full line
and shows a clear $\cos \Phi$ dependence with a small $\cos2\Phi$ modulation.
The extracted values for the combination $(2 \rcuu + \rczz)$ are 
presented in Fig.~\ref{fig:phi2}b for three bins in \qsq. The
\rh\ and \om\ background subtraction in the $\Phi$ distribution reduces the
value of the combination $(2 \rcuu + \rczz)$ by 13 \% (around half
of the statistical error). In Fig.~\ref{fig:phi2}b, the effect of 
QED radiative corrections on the measurement of the combination $(2 \rcuu + \rczz)$ 
was taken into account. This effect was estimated using the DIFFVM simulation including
a HERACLES~\cite{heracles} interface, 
and reduces the observed value of the combination $(2 \rcuu + \rczz)$
by 17~\%. The combination $(2 \rcuu + \rczz)$ obtained from the fit
deviates significantly from the zero prediction of SCHC (a 5~$\sigma$ effect).
The values of the combination $(2 \rcuu + \rczz)$ are similar to the ones
obtained in case of elastic \rh\ meson production~\cite{h1_rho_96}.

From the measurement of the spin density matrix elements, the ratio $R$ of
cross sections for \ph\ meson production by longitudinal and transverse 
virtual photons can be extracted.
As the SCHC violating amplitudes are small compared to the 
helicity conserving amplitudes, one can make\footnote{
 The effect of SCHC violation on the measurement of $R$ is 
 of the order of 3 \%.} the
SCHC approximation in order to estimate $R$, which is then obtained directly from the 
measurement of the matrix element \rzqzz~\cite{h1_rho_96}.

\begin{table}[t]
\begin{center}
\begin{tabular}{|c|c|}
\hline
$Q^2$ (GeV$^2$)   &  $R = \sigma_L / \sigma_T$   \\
\hline
2.0  & 0.47  \mbig{$^{+0.26}_{-0.19}$}  \mbig{$^{+0.07}_{-0.06}$}     \\
2.9  & 0.87  \mbig{$^{+0.38}_{-0.27}$}  \mbig{$^{+0.20}_{-0.06}$}    \\
4.5  & 1.48  \mbig{$^{+0.82}_{-0.49}$}  \mbig{$^{+0.52}_{-0.09}$}    \\
8.6  & 5.9  \mbig{$^{+5.6}_{-2.1}$}  \mbig{$^{+1.8}_{-0.5}$}    \\
\hline
\end{tabular}
\end{center}
  \caption{Ratio of the longitudinal to transverse cross sections 
    for elastic \ph\ meson production, for four \qsq\ values. 
    The first error represents the statistical error and the second the
    systematic error.}
\label{table:r}
\end{table}

\begin{figure}[p]
\begin{center}
\setlength{\unitlength}{1.0cm}
\begin{picture}(16.0,8.0)
\put(-0.4,0.0){\epsfig{file=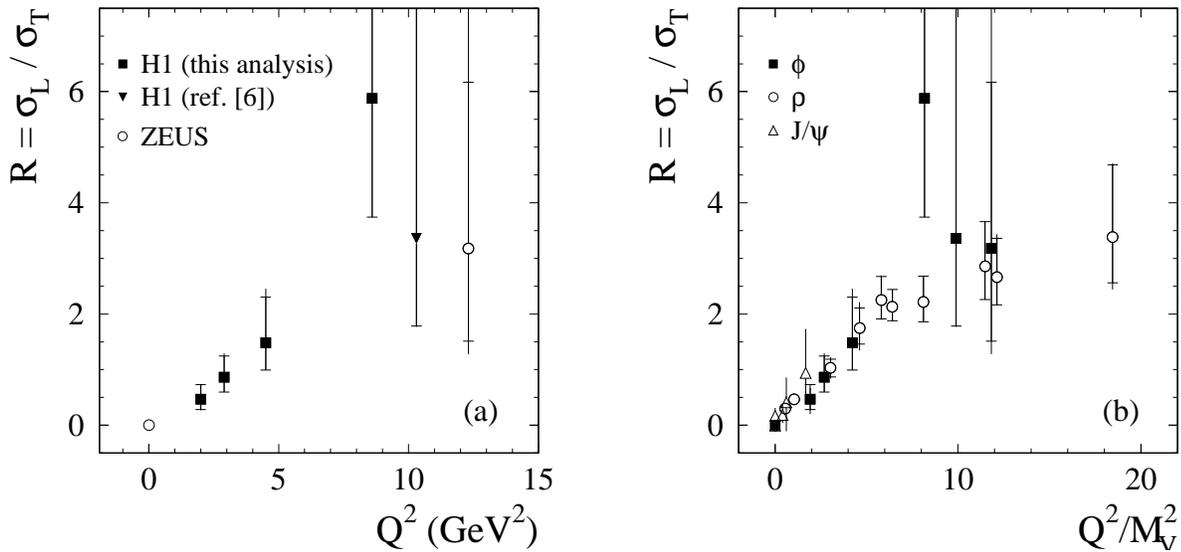}}%
\end{picture}
\caption{Ratio of the longitudinal to transverse cross sections 
    for elastic \ph\ meson production, as a function of \qsq\ 
    ~\protect\cite{h1_phi_94,zeus_phi_hq,zeus_phi_gp} (a) and for various
    vector mesons as 
    a function of $Q^2/M_V^2$ (b)~\protect\cite{h1_rho_96,
    zeus_rho_jpsi_hq,h1_jpsi_gp,h1_jpsi_hq,zeus_jpsi_gp,
    h1_phi_94,zeus_phi_gp,zeus_phi_hq,zeus_rho_gp,zeus_ome_gp}.
    The inner error bars are statistical and the full
    error bars include the systematic errors added in quadrature.}
\label{fig:r}
\end{center}
\end{figure}

The \qsq\ dependence of $R$ is presented in Fig.~\ref{fig:r}a,
together with other measurements performed under the SCHC approximation
~\cite{h1_phi_94,zeus_phi_hq,zeus_phi_gp}, see also table~\ref{table:r}.
It is observed that $R$ rises steeply with \qsq, and that the longitudinal
cross section dominates over the transverse cross section for
$Q^2 \gsim$ 3~\gevsq. 
The rise of $R$ with \qsq\ for \ph\ meson production is slower than 
for the \rh\ meson~\cite{h1_rho_96}. However, when plotted as a function of
$Q^2/M_V^2$, the ratio $R$
appears to show a common dependence for different vector
mesons~\cite{h1_rho_96,zeus_rho_jpsi_hq,h1_jpsi_gp,h1_jpsi_hq,zeus_jpsi_gp,
h1_phi_94,zeus_phi_gp,zeus_phi_hq,zeus_rho_gp,zeus_ome_gp},
see Fig.~\ref{fig:r}b (for further details see ref.~\cite{bar_tel_aviv}). 
%
%

%
\section{Summary} 
The elastic electroproduction of \ph\ mesons has been studied 
with the H1 detector in the kinematic range 1$<$ \qsq\ $<$ 15~\gevsq\
and 40 $<$ \W\ $<$ 130 GeV.
The \qsq\ dependence of the cross section is presented in the form of
the ratio to the elastic \rh\ meson cross section. A significant rise of the ratio
with \qsq\ is observed. The elastic \ph\ meson cross section is 
extracted using recent H1 results of elastic \rh\ meson production. 
A compilation of the elastic \rh, \om, \ph,
\jpsi\ and $\Upsilon$ meson cross sections, scaled by SU(5) factors,
is presented as a function of (\qsq\ + $M_V^2$).
A common dependence is observed within experimental errors. 
The \modt\ dependence of the elastic \ph\ meson cross section 
is well described by an exponentially falling distribution. 
The full set of spin density matrix elements is measured in 
two \qsq\ bins. Predictions based on perturbative QCD are 
compared to the measurements.
The combination $(2 \rcuu\ + \rczz)$ is extracted from the $\Phi$ angle
distribution and is observed to deviate from zero, which indicates
a small but significant violation of the $s$-channel 
helicity conservation (SCHC) approximation.
The ratio $R$ of longitudinal to transverse \ph\ meson production cross sections
is observed to increase with \qsq. 
A common dependence for $R$ as a function of $Q^2/M_V^2$
is observed for elastic \rh, \ph\ and \jpsi\ meson production.

\section*{Acknowledgements}
We are grateful to the HERA machine group whose outstanding
efforts have made and continue to make this experiment possible.
We thank
the engineers and technicians for their work in constructing and now
maintaining the H1 detector, our funding agencies for
financial support, the
DESY technical staff for continual assistance,
and the DESY directorate for the
hospitality which they extend to the non DESY
members of the collaboration.
We thank further I.~Akushevich, J.-R.~Cudell, D.Yu.~Ivanov, N.~Nikolaev
and I.~Royen for useful 
discussions and for providing us with their model predictions.


%
\end{document}